\documentclass[aps,twocolumn,superscriptaddress,showpacs]{revtex4}
\usepackage{amsfonts}
\usepackage{amsmath}
\usepackage{amssymb}
\usepackage{graphicx}

\begin{document}

\title{Quantum Phases of Bose-Hubbard Model in Optical Superlattices}
\author{Bo-Lun Chen}
\affiliation{Department of Physics, Beijing Normal University, Beijing 100875, P. R. China}
\author{Su-Peng Kou}
\email{spkou@bnu.edu.cn}
\affiliation{Department of Physics, Beijing Normal University, Beijing 100875, P. R. China}
\author{Yunbo Zhang}
\affiliation{Institute of Theoretical Physics, Shanxi University, Taiyuan 030006, P. R.
China}
\author{Shu Chen}
\affiliation{Institute of Physics, Chinese Academy of Sciences, Beijing 100190, P. R.
China}

\begin{abstract}
In this paper, we analyze the quantum phases of multiple component
Bose-Hubbard model in optical superlattices, using a mean-field
method, the decoupling approximation. We find that the phase
diagrams exhibit complected patterns and regions with various Charge
Density Wave (CDW) for both one- and two- component cases. We also
analyze the effective spin dynamics for the two-component case in
strong-coupling region at unit filling, and show the possible
existence of a Spin Density Wave (SDW) order.
\end{abstract}

\maketitle

\section{Introduction}

The observation of Mott insulator -- superfluid transition of ultracold
bosons loaded in optical lattices\cite{Greiner} has triggered huge amount of
interest in quantum simulation\cite{Buluta}. A lot of efforts have been made
to investigate Bose-Hubbard Hamiltonian\cite{Fisher}, both single\cite{Stoof}
and multiple components\cite{chen}, in cold atom systems (in double-wells\cite%
{Zhang}, superlattices\cite{Buonsante1,Barthel}), using various techniques
(projection wave-function\cite{Zwerger}, decoupling approximation\cite%
{Stoof,Chen}, field theory\cite{Dupuis}, dynamical mean-field\cite{Amico},
etc.), aiming to achieve a comprehensible understanding of this many-body
model\cite{Yukalov}.

Among these, multi-component Bose-Hubbard in optical superlattices attracts
special attention due to its rich quantum phases\cite{Duan,Altman} and
promising potential in emulating effective spin-spin interactions\cite%
{Barthel,Bloch}. Many works on this topic have revealed the existence of
Mott-insulator with integer and fractional fillings, which correspond to
various Charge Density Wave (CDW) and complected dynamical behaviors\cite%
{Kuklov,Ziegler,Roth,Cirac,Rousseau,Bhattacherjee}. People also reported
topological phases and exotic localizations in disordered superlattices\cite%
{Buonsante2,Pollet}. Recently, to obtain a complete and thorough description
of this model, several attempts have been made\cite{Iskin,Hen}.

In this paper, we reinvestigate this problem by using a mean-field approach,
the decoupling approximation. Working in the Mott insulating phase as an
unperturbed ground state, and treating the hopping as perturbations, this
method (though at mean-field level) can still give comparatively accurate
results for phase transition, comparing with numerical simulations\cite%
{Stoof}. Besides, since this method decouples the original Hamiltonian into
a set of single-site Hamiltonian, it exhibits simple energy\ expressions for
different filling situations in the Mott state. Therefore, we can analyze
and write down all the possible filling configurations in a clear and
evident manner. Then, We depict the corresponding phase boundaries for each
filling condition, and gain a systematic description of possible quantum
phases of this model. We further consider the effective spin dynamics at
strong coupling limit and unit filling, and find both a ferromagnetic and an
anti-ferromagnetic\ spin-wave excitation as the potential barrier between
neighboring sites varies.

The paper is organized as follow: We first introduce the model in Sec II,
then deals with the single-component Bose-Hubbard model in Sec. III, which
acts as a precedent of the two-component case that is discussed in Sec. IV.
Both sections contain the mean-field calculations, the analysis of possible
filling patterns and phase diagrams with corresponding illustrations. The
spin dynamics is analyzed in the last subsection in Sec. IV. We give the
conclusion in Sec. V.

\section{Models}

Our starting point is a two-component Bose-Hubbard model in a
double-periodic superlattices,
\begin{eqnarray}
\hat{H} &=&-\sum_{\left\langle ij\right\rangle \sigma }(t_{ij}\hat{a}%
_{i\sigma }^{\dag }\hat{a}_{j\sigma }+h.c.)+\frac{U}{2}\sum_{i\sigma }\hat{n}%
_{i\sigma }\left( \hat{n}_{i\sigma }-1\right)  \notag \\
&&+V\sum_{i}\hat{n}_{i\uparrow }\hat{n}_{i\downarrow }-\sum_{i\sigma }\left(
\mu _{\sigma }-\Delta _{i}\right) \hat{n}_{i\sigma }.  \label{2c BH}
\end{eqnarray}%
Here $\left\langle ij\right\rangle $ denotes the nearest-neighbor counting, $%
t_{ij}$ is the hopping amplitude, $U$ is the inter-species repulsion, $V$ is
the intra-species repulsion, $\mu _{\sigma }$ is the chemical potential that
restricts the particle number, $\Delta _{i}$ is the energy bias on a given
site, $\hat{n}_{i\sigma }=\hat{a}_{i\sigma }^{\dag }\hat{a}_{i\sigma }$ is
the number operator for bosons with $\sigma \equiv \uparrow ,\downarrow $
representing the two internal states of the trapped Bose Einstein Condensate
(BEC).

For single-component Bose-Hubbard, the (pseudo)spin gets polarized so that $%
\sigma $ takes a certain value; thus there is no intra-species interaction $%
V $.

\begin{figure}[tbp]
\begin{center}
\includegraphics[width=0.4\textwidth]{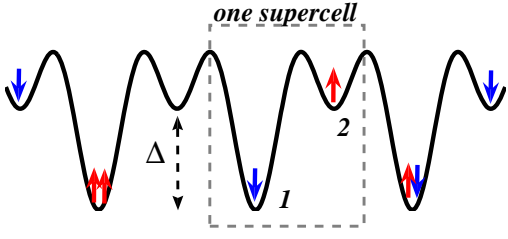}
\end{center}
\caption{(color online) An illustration of a period-$2$ superlattice. The
potential barrier is $\Delta $. $1\left( 2\right) $ is denoted as the deep
(shallow) site in a supercell. The red and blue arrows represent different
internal states of the trapped BEC. }
\label{superlattice}
\end{figure}

Generally speaking, the period of the superlattice can be set to be an
arbitrary integer $l$, as long as we require physical quantities are
periodic functions, $\psi _{i}=\psi _{i+l}$. For $l=2$, we can simply set
the potential bias to be
\begin{equation}
\Delta _{i}=\left\{
\begin{array}{c}
0,i\in \text{\textrm{odd}} \\
\Delta ,i\in \text{\textrm{even}}%
\end{array}%
\right. ,  \label{biased potential}
\end{equation}%
as shown in Fig. (\ref{superlattice}).

\section{One-component case}

\subsection{Basic Formulism}

First, let us consider the one-component case for clarity. The Hamiltonian
reduces to%
\begin{eqnarray}
\hat{H} &=&-\sum_{\left\langle ij\right\rangle }(t_{ij}\hat{a}_{i}^{\dag }%
\hat{a}_{j}+h.c.)+\frac{U}{2}\sum_{i}\hat{n}_{i}\left( \hat{n}_{i}-1\right)
\notag \\
&&-\sum_{i}\left( \mu -\Delta _{i}\right) \hat{n}_{i}.  \label{1c BH}
\end{eqnarray}%
To determine the phase boundary of Mott-insulator (MI) -- superfluid (SF)
transition, we apply an extended decoupling approximation which was
developed in Ref. \cite{Stoof}. In this mean-field scheme, we write the two
bosonic operators $\hat{a}_{i}^{\dag }\hat{a}_{j}$ as $\hat{a}_{i}^{\dag }%
\hat{a}_{j}\approx \langle \hat{a}_{i}^{\dag }\rangle \hat{a}_{j}+\hat{a}%
_{i}^{\dag }\langle \hat{a}_{j}\rangle -\langle \hat{a}_{i}^{\dag }\rangle
\langle \hat{a}_{j}\rangle $. Then by introducing a site-dependent (local
and small) SF order parameter, $\psi _{i}\equiv \langle \hat{a}_{i}^{\dag
}\rangle =\langle \hat{a}_{i}\rangle $, the hopping term can be written as,
\begin{eqnarray}
-\sum_{\left\langle ij\right\rangle }(t_{ij}\hat{a}_{i}^{\dag }\hat{a}%
_{j}+h.c.) &\approx &-t\sum_{i,\mathbf{\delta }}[\psi _{i}(\hat{a}_{i+%
\mathbf{\delta }}+\hat{a}_{i+\mathbf{\delta }}^{\dag })-\psi _{i}\psi _{i+%
\mathbf{\delta }}  \notag \\
&&\ \ \ \ \ \ \ \ +\psi _{i+\mathbf{\delta }}(\hat{a}_{i}+\hat{a}_{i}^{\dag
})-\psi _{i+\mathbf{\delta }}\psi _{i}]  \notag \\
&=&-2t\sum_{i}(\hat{a}_{i}^{\dag }+\hat{a}_{i}-\psi _{i})\sum_{\mathbf{%
\delta }}\psi _{i+\mathbf{\delta }}  \notag \\
&=&-2zt\sum_{i}(\hat{a}_{i}^{\dag }+\hat{a}_{i}-\psi _{i})\psi _{i+1},
\label{1c decouple}
\end{eqnarray}%
where we assume that in the strong-coupling regime ($U,V\gg t_{ij},t_{ji}$) $%
t_{ij}\approx t_{ji}\equiv t$ and denote $\psi _{i+\mathbf{\delta }}=\psi
_{i-\mathbf{\delta }}\equiv \psi _{i+1}$, $\mathbf{\delta }$ is the position
vector for the nearest neighbors, $z$ is the coordination number. We then
decouple the initial Hamiltonian Eq. (\ref{1c BH})\ into $N$ (nearly)
independent parts\cite{Buonsante1}, $\hat{H}=\sum_{i=1}^{N}\hat{H}_{i}$,
where
\begin{eqnarray}
\hat{H}_{i} &=&\frac{U}{2}\hat{n}_{i}\left( \hat{n}_{i}-1\right) -\left( \mu
-\Delta _{i}\right) \hat{n}_{i}  \notag \\
&&-2zt(\hat{a}_{i}^{\dag }+\hat{a}_{i}-\psi _{i})\psi _{i+1}.
\end{eqnarray}%
As a result, we can concentrate in one super-cell as shown in Fig. (\ref%
{superlattice}) with reduced Hamiltonian
\begin{eqnarray}
\bar{H}_{sc} &\equiv &\left( 2zt\right) ^{-1}\sum_{l=1}^{2}\hat{H}_{l}
\notag \\
&=&\sum_{l=1}^{2}[\frac{\bar{U}}{2}\hat{n}_{l}\left( \hat{n}_{l}-1\right)
-\left( \bar{\mu}-\bar{\Delta}_{l}\right) \hat{n}_{l}  \notag \\
&&\ \ \ \ -(\hat{a}_{l}^{\dag }+\hat{a}_{l}-\psi _{l})\psi _{l+1}]  \notag \\
&=&\bar{H}_{sc}^{\left( 0\right) }+\psi _{2}\hat{V}_{1}+\psi _{1}\hat{V}_{2},
\label{1c H_sc}
\end{eqnarray}%
where the dimensionless quantities are $\bar{U}\equiv U/2zt$, $\bar{\mu}%
\equiv \mu /2zt$ and $\bar{\Delta}_{l}\equiv \Delta _{l}/2zt$, the
unperturbed Hamiltonian $\bar{H}_{sc}^{\left( 0\right) }$ is
\begin{equation}
\bar{H}_{sc}^{\left( 0\right) }=\sum_{l=1}^{2}[\frac{\bar{U}}{2}\hat{n}%
_{l}\left( \hat{n}_{l}-1\right) -\bar{\mu}_{l}\hat{n}_{l}]+2\psi _{1}\psi
_{2}  \label{1c H0}
\end{equation}%
with reduced chemical potential $\bar{\mu}_{l}\equiv \bar{\mu}-\bar{\Delta}%
_{l}$, while the hopping-induced perturbations are
\begin{equation}
\hat{V}_{l}=-(\hat{a}_{l}^{\dag }+\hat{a}_{l}).  \label{1c V}
\end{equation}%
In deriving above equations, we have used the periodic condition of the
superlattice, $\psi _{l+2}=\psi _{l}$, $l=1,2$.

The unperturbed energy is the energy of states that have definite particle
number (Mott state), namely,
\begin{equation}
E_{g}^{\left( 0\right) }\equiv E_{\left\{ g_{1};g_{2}\right\} }^{\left(
0\right) }=\min \{E_{n}^{\left( 0\right) }\}_{n=0,1,2,\ldots },
\end{equation}%
where $g=g_{1}+g_{2}$ is the average particle number in one supercell, $%
g_{l} $ is the particle number on a certain site $l$ in the supercell. Up to
the second order, this implies
\begin{equation}
E_{g}^{\left( 0\right) }<E_{g+1}^{\left( 0\right) },\ E_{g}^{\left( 0\right)
}<E_{g-1}^{\left( 0\right) },  \label{energy comparison}
\end{equation}%
thus there will be a constraint on the chemical potential $\bar{\mu}$. (To
be discussed later.)

In the Mott state and near the MI-SF transition, the on-site particle number
is still well-defined, then the unperturbed energy (written in a
dimensionless form) becomes%
\begin{eqnarray}
E_{\left\{ g_{1};g_{2}\right\} }^{\left( 0\right) } &=&\sum_{l=1}^{2}[\frac{%
\bar{U}}{2}g_{l}\left( g_{l}-1\right) -\bar{\mu}_{l}g_{l}]+2\psi _{1}\psi
_{2}  \notag \\
&=&\frac{\bar{U}}{2}\left( g_{1}^{2}+g_{2}^{2}-g\right) -\bar{\mu}g+\bar{%
\Delta}g_{2}+2\psi _{1}\psi _{2}  \label{1c E0_1}
\end{eqnarray}%
where we have used Eq. (\ref{biased potential}).

Meanwhile, the perturbation term $\hat{V}$ results in an energy correction,
we can calculate this second-order perturbation energy $E_{g}^{\left(
2\right) }$ in a standard manner \cite{Stoof},
\begin{eqnarray}
E_{g}^{\left( 2\right) } &\equiv &E_{\left\{ g_{1};g_{2}\right\} }^{\left(
2\right) }  \notag \\
&=&\psi _{2}^{2}\sum_{n_{1}+n_{2}\neq g_{1}+g_{2}}\frac{|\langle n_{1};n_{2}|%
\hat{V}_{1}|g_{1};g_{2}\rangle |^{2}}{E_{\left\{ g_{1};g_{2}\right\}
}^{\left( 0\right) }-E_{\left\{ n_{1};n_{2}\right\} }^{\left( 0\right) }}+
\notag \\
&&\psi _{1}^{2}\sum_{n_{1}+n_{2}\neq g_{1}+g_{2}}\frac{|\langle n_{1};n_{2}|%
\hat{V}_{2}|g_{1};g_{2}\rangle |^{2}}{E_{\left\{ g_{1};g_{2}\right\}
}^{\left( 0\right) }-E_{\left\{ n_{1};n_{2}\right\} }^{\left( 0\right) }},
\end{eqnarray}%
where the summation over $\left\{ n_{1};n_{2}\right\} $ has only the
following terms, $\left\{ n_{1};n_{2}\right\} =\left\{ g_{1}\pm
1;g_{2}\right\} ,\,\left\{ g_{1};g_{2}\pm 1\right\} $, due to the simple
form of $\hat{V}_{l}$. After straightforward calculations, we find
\begin{eqnarray}
E_{\left\{ g_{1};g_{2}\right\} }^{\left( 2\right) } &=&\psi _{2}^{2}\left[
\frac{g_{1}+1}{-\bar{U}g_{1}+\bar{\mu}}+\frac{g_{1}}{\bar{U}\left(
g_{1}-1\right) -\bar{\mu}}\right] +  \notag \\
&&\psi _{1}^{2}\left[ \frac{g_{2}+1}{-\bar{U}g_{2}+\bar{\mu}-\bar{\Delta}}+%
\frac{g_{2}}{\bar{U}\left( g_{2}-1\right) -\bar{\mu}+\bar{\Delta}}\right]
\notag \\
&=&\frac{-\psi _{2}^{2}\left( \bar{\mu}+\bar{U}\right) }{\left[ \bar{U}%
\left( g_{1}-1\right) -\bar{\mu}\right] \left( -\bar{U}g_{1}+\bar{\mu}%
\right) }+  \notag \\
&&\frac{-\psi _{1}^{2}\left( \bar{\mu}-\bar{\Delta}+\bar{U}\right) }{\left[
\bar{U}\left( g_{2}-1\right) -\bar{\mu}+\bar{\Delta}\right] \left( -\bar{U}%
g_{2}+\bar{\mu}-\bar{\Delta}\right) }.  \label{1c E2}
\end{eqnarray}%
Therefore, the total energy of a super-cell is $E=E_{g}^{\left( 0\right)
}+E_{g}^{\left( 2\right) }+\cdots $ can be expanded into a power series of
the SF order parameter $\psi _{1}$ and $\psi _{2}$ (Landau expansion),
\begin{equation}
E\left( \psi _{1},\psi _{2}\right) =a_{0}+a_{2}\psi _{1}^{2}+b_{2}\psi
_{2}^{2}+c_{2}\psi _{1}\psi _{2}+\mathcal{O}\left( \psi _{1}^{4},\psi
_{2}^{4}\right) ,  \label{1c Landau expansion}
\end{equation}%
where we presume all coefficients of fourth-order and above are positive,
stabilizing the system. The coefficients are

\begin{subequations}
\label{1c_coeffient}
\begin{eqnarray}
a_{0} &=&\frac{\bar{U}}{2}\left( g_{1}^{2}+g_{2}^{2}-g\right) -\bar{\mu}g+%
\bar{\Delta}g_{2}, \\
a_{2} &=&\frac{-\left( \bar{\mu}-\bar{\Delta}+\bar{U}\right) }{\left[ \bar{U}%
\left( g_{2}-1\right) -\bar{\mu}+\bar{\Delta}\right] \left( -\bar{U}g_{2}+%
\bar{\mu}-\bar{\Delta}\right) }, \\
b_{2} &=&\frac{-\left( \bar{\mu}+\bar{U}\right) }{\left[ \bar{U}\left(
g_{1}-1\right) -\bar{\mu}\right] \left( -\bar{U}g_{1}+\bar{\mu}\right) },\
c_{2}=2.
\end{eqnarray}
\end{subequations}

Clearly, $E\left( \psi _{1}=0,\psi _{2}=0\right) $ is a local
extremum, and
represents the MI state. As the interaction $\bar{U}$, the biased potential $%
\bar{\Delta}$ and the particle number (chemical potential $\bar{\mu}$)
changes to a critical point, the extremum $E\left( 0,0\right) $ becomes an
instability point, thus a\ phase transition to SF state occurs. To express
this idea explicitly, we should compare $\partial _{\psi _{1}}^{2}E|_{\left(
0,0\right) }$, $\partial _{\psi _{2}}^{2}E|_{\left( 0,0\right) }$ and $%
\partial _{\psi _{1}}\partial _{\psi _{2}}E|_{\left( 0,0\right) }$. The
critical condition is $\partial _{\psi _{1}}^{2}E|_{\left( 0,0\right)
}\partial _{\psi _{2}}^{2}E|_{\left( 0,0\right) }=\left( \partial _{\psi
_{1}}\partial _{\psi _{2}}E|_{\left( 0,0\right) }\right) ^{2}$, i.e., $%
4a_{2}b_{2}=c_{2}^{2}$, or equivalently,
\begin{widetext}
\begin{equation}
1=\frac{\left( \bar{\mu}+\bar{U}\right) \left( \bar{\mu}-\bar{\Delta}+\bar{U}%
\right) }{\left[ \bar{U}\left( g_{1}-1\right) -\bar{\mu}\right] \left( -\bar{%
U}g_{1}+\bar{\mu}\right) \left[ \bar{U}\left( g_{2}-1\right) -\bar{\mu}+\bar{%
\Delta}\right] \left( -\bar{U}g_{2}+\bar{\mu}-\bar{\Delta}\right) }.
\end{equation}%
Simplify it, we obtain the phase boundary,
\begin{equation}
\tilde{t}_{c}=\frac{1}{2}\sqrt{\frac{\left[ \left( g_{1}-1\right) -\tilde{\mu%
}\right] \left( -g_{1}+\tilde{\mu}\right) [\left( g_{2}-1\right) -\tilde{\mu}%
+\tilde{\Delta}](-g_{2}+\tilde{\mu}-\tilde{\Delta})}{\left( \tilde{\mu}%
+1\right) (\tilde{\mu}-\tilde{\Delta}+1)}},  \label{1c t_c}
\end{equation}%
\end{widetext}where $\tilde{t}_{c}\equiv zt/U$, $\tilde{\mu}\equiv \mu /U$,
and $\tilde{\Delta}\equiv \Delta /U$.

Before we come to the phase diagram, let us make some remarks on our
approximation method. As pointed out in Ref. \cite{Chen}, the
decoupling scheme works well in the vicinity of Mott state. The
hopping effect induces a weak and local superfluid $\psi _{i}$,
serving as a perturbation over the Mott state. Consequently,
although this technique can determine the phase boundary
conveniently, it cannot extrapolate correct physics in deep
superfluid phase. Besides, from Eqs. (\ref{1c H_sc}, \ref{1c H0}, \ref{1c V}%
) and the introduction of many dimensionless quantities we know that the
MI-SF transition is universal, i.e., it occurs in similar manners in
different dimensions; and this approximation gets better accuracy in higher
dimensions.

\subsection{Phase Diagrams}

To depict the phase diagram, we can plot the phase boundary surface, Eq. (%
\ref{1c t_c}) in the $\tilde{\mu}$-$\tilde{\Delta}$-$\tilde{t}$ coordinate
system. Before that, we need to take a closer look at the unperturbed
groundstate to find appropriate constraints on the chemical potential $%
\tilde{\mu}$.

As we mentioned, assuming the average filling in a supercell is $%
g=g_{1}+g_{2}\equiv \left\{ g_{1};g_{2}\right\} $, a natural question
arises: what is the most energetically favorable filling factor $\left\{
g_{1};g_{2}\right\} $ for a given $g$? A straightforward solution to this
question is to compare two arbitrary filling configurations $\left\{
k_{1};k_{2}\right\} $ and $\left\{ m_{1};m_{2}\right\} $ with $%
k_{1}+k_{2}=m_{1}+m_{2}=g$; namely, to compare their unperturbed energy, $%
\Delta E=E_{\left\{ k_{1};k_{2}\right\} }^{\left( 0\right) }-E_{\left\{
m_{1};m_{2}\right\} }^{\left( 0\right) }$. By using Eq. (\ref{1c E0_1}),
this energy difference is

\begin{equation}
\widetilde{\Delta E}=\left( k_{2}+m_{2}-g\right) \left( k_{2}-m_{2}\right)
+\left( k_{2}-m_{2}\right) \tilde{\Delta}  \label{1c energy diff}
\end{equation}%
with $\widetilde{\Delta E}\equiv \Delta E/U$, $k_{2},m_{2}=0,1,\ldots ,g$,
respectively. Therefore, for a given potential bias $\tilde{\Delta}\equiv
\Delta /U$ and filling number $g$, we can determine the groundstate
configuration $\left\{ g_{1};g_{2}\right\} $ which has the largest
differences $\widetilde{\Delta E}$ with respect to all other configurations.

More intuitively, we can simply draw the energy levels in the two sites of a
supercell with a given reduced potential bias $\tilde{\Delta}$, and then
fill atoms from the lowest level to higher levels one by one, up to $g$.
This filling sequence naturally costs the least energy, thus the resulting
configuration is just the groundstate filling $\left\{ g_{1};g_{2}\right\} $
(see Appendix).

As a result, for a given groundstate $\left\{ g_{1};g_{2}\right\} $, from
Eq. (\ref{energy comparison}) we obtain the constraint on chemical potential
$\tilde{\mu}$. For the most imbalanced case $\left\{ g;0\right\} ,$ $\tilde{%
\mu}\in \left( g-1,g\right) $. For other cases, the calculations are simple
but tedious. The results are listed in the following table. ($p\in \mathrm{%
even}$.)
\begin{widetext}
\begin{equation*}
\begin{tabular}{|c|c|c|}
\hline $\tilde{\Delta}$ & $g\in \mathrm{even}$ & $g\in \mathrm{odd}$
\\ \hline $\left( p-1,p\right) $ & $\bar{\mu}\in (\frac{1}{2}\left[
g+\left(
p-2\right) \right] ,\frac{1}{2}\left( g-p\right) +\tilde{\Delta})$ & $\bar{%
\mu}\in (\frac{1}{2}\left[ g-\left( p+1\right) \right] +\tilde{\Delta},\frac{%
1}{2}\left[ g+\left( p-1\right) \right] )$ \\ \hline
$\left( p,p+1\right) $ & $\bar{\mu}\in (\frac{1}{2}\left[ g-\left(
p+2\right) \right] +\tilde{\Delta},\frac{1}{2}\left( g+p\right) )$ & $\bar{%
\mu}\in (\frac{1}{2}\left[ g+\left( p-1\right) \right] ,\frac{1}{2}\left[
g-\left( p+1\right) \right] +\tilde{\Delta})$ \\ \hline
\end{tabular}%
\end{equation*}%
\end{widetext}

\begin{figure}[tbp]
\includegraphics[width=0.5\textwidth]{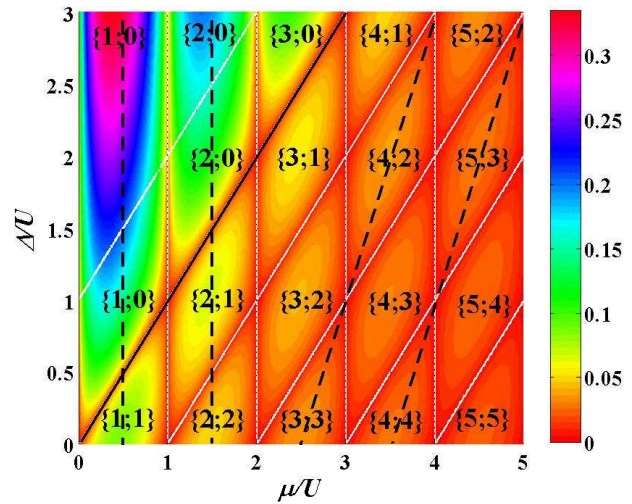}
\caption{(color online) A contour diagram for one-component case. Regions
with different numbers correspond to different unperturbed ground state
filling patterns. The left (right) number represents the particle number in
deep (shallow) sit in a supercell [see Fig. (\protect\ref{superlattice})].
This filling configuration reflects a CDW order in the Mott state. Regions
to the left of the black solid line favor an imbalanced filling pattern $%
\left\{ g;0\right\} $. The chemical potential $\protect\mu $ (particle
number $g$) is fixed along the vertical (oblique) black dashed lines. }
\label{1c_phase_diagram_2D}
\end{figure}

With these restrictions in hand, we now plot the phase diagram for
one-component Bose-Hubbard model in superlattices, as shown in Figs. (\ref%
{1c_phase_diagram_3D}) and (\ref{1c_phase_diagram_2D}). It can be seen that
in strong-coupling regions ($t\ll U$, under the lobes) with given potential
bias $\Delta $, the system is in Mott phase. As $U$ decreases, phase
transitions occur, the system goes into superfluid phase that is above the
lobes. On the other hand, when $\Delta $ changes, the phase boundaries and
allowed Mott states change correspondingly.

\begin{figure}[tbp]
\includegraphics[width=0.5\textwidth]{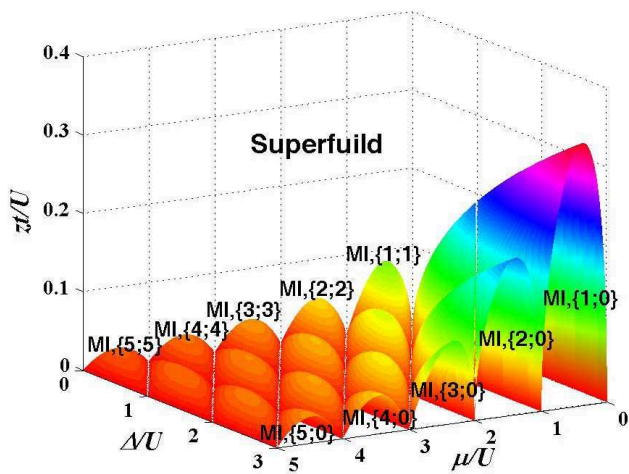}
\caption{(color online) The phase diagram for one-component case. Regions
under the lopes are Mott insulating phase (MI), regions above those lopes
are superfluid phase. }
\label{1c_phase_diagram_3D}
\end{figure}

\begin{figure}[tbp]
\begin{center}
\includegraphics[width=0.52\textwidth]{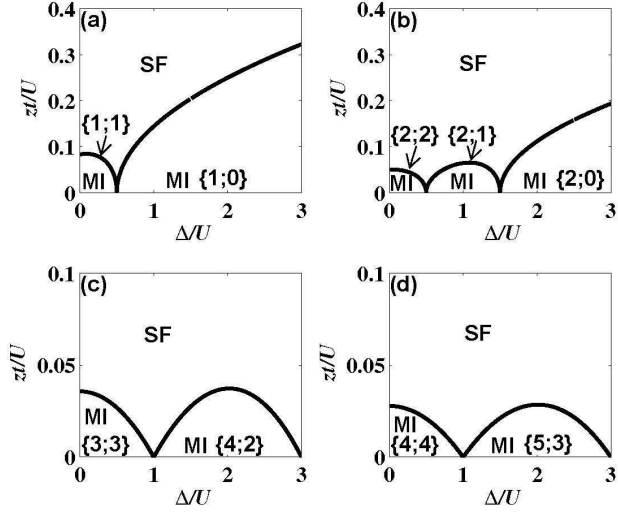}
\end{center}
\caption{The phase boundaries along the dashed lines in Fig. (\protect\ref%
{1c_phase_diagram_2D}). In (a) and (b) chemical potential $\protect\mu $ is
fixed, in (c) and (d) particle number $g$ is fixed. Notations are defined in
Figs. (\protect\ref{1c_phase_diagram_3D}) and (\protect\ref%
{1c_phase_diagram_2D}). SF stands for superfluid phase.}
\label{1c_phase_diagram_crosssections}
\end{figure}

For clarity, we denote the groundstate filling configuration at different
parameter regions in Fig. (\ref{1c_phase_diagram_2D}). In regions to the
left of the black solid line, parameters are chosen such that in Mott phase
particles in a supercell must reside in the deep site (site $1$), thus forms
crystalline structure that has a doubled period of the original optical
lattice. To illustrate the effect of increasing potential bias, we draw the
phase boundaries along the vertical black dashed lines with fixed chemical
potential $\mu $, and along the oblique black dashed lines with fixed
particle numbers $g$ in Figs. (\ref{1c_phase_diagram_crosssections}).

\section{Two-component case}

\subsection{Basic formulism}

The Hamiltonian is just Eq. (\ref{2c BH}). The hopping term can be similarly
decoupled into a summation over single-site parts, $\hat{a}_{i\sigma }^{\dag
}\hat{a}_{j\sigma }=\langle \hat{a}_{i\sigma }^{\dag }\rangle \hat{a}%
_{j\sigma }+\hat{a}_{i\sigma }^{\dag }\langle \hat{a}_{j\sigma }\rangle
-\langle \hat{a}_{i\sigma }^{\dag }\rangle \langle \hat{a}_{j\sigma }\rangle
$, assuming the SF order parameter is also spin-dependent, $\psi _{i\sigma
}\equiv \langle \hat{a}_{i\sigma }^{\dag }\rangle =\left\langle \hat{a}%
_{i\sigma }\right\rangle $. Thus, like the one-component case, the hopping
term becomes%
\begin{equation}
-t\sum_{\left\langle ij\right\rangle \sigma }(\hat{a}_{i\sigma }^{\dag }\hat{%
a}_{j\sigma }+h.c.)\approx -2zt\sum_{i\sigma }(\hat{a}_{i\sigma }^{\dag }+%
\hat{a}_{i\sigma }-\psi _{i\sigma })\psi _{i+1,\sigma },
\end{equation}%
and decoupled Hamiltonian is
\begin{eqnarray}
\hat{H}_{i} &=&\frac{U}{2}\sum_{\sigma }\hat{n}_{i\sigma }\left( \hat{n}%
_{i\sigma }-1\right) +V\hat{n}_{i\uparrow }\hat{n}_{i\downarrow
}-\sum_{\sigma }\left( \mu _{\sigma }-\Delta _{i}\right) \hat{n}_{i\sigma }
\notag \\
&&-2zt\sum_{\sigma }(\hat{a}_{i\sigma }^{\dag }+\hat{a}_{i\sigma }-\psi
_{i\sigma })\psi _{i+1,\sigma }.
\end{eqnarray}%
The effective Hamiltonian in one supercell $\bar{H}_{sc}$ can still be
written into two parts, $\bar{H}_{sc}=\left( 2zt\right) ^{-1}\sum_{l=1}^{2}%
\hat{H}_{l}=\hat{H}_{sc}^{\left( 0\right) }+\hat{V}$, where the unperturbed
part is
\begin{eqnarray}
\hat{H}_{sc}^{\left( 0\right) } &=&\sum_{l=1,\sigma }^{2}[\frac{\bar{U}}{2}%
\hat{n}_{l\sigma }\left( \hat{n}_{l\sigma }-1\right) -\bar{\mu}_{l\sigma }%
\hat{n}_{l\sigma }+\psi _{l\sigma }\psi _{l+1,\sigma }]  \notag \\
&&+\bar{V}\sum_{l=1}^{2}\hat{n}_{l\uparrow }\hat{n}_{l\downarrow }
\label{2c H_0}
\end{eqnarray}%
with $\bar{\mu}_{l\sigma }\equiv \bar{\mu}_{\sigma }-\bar{\Delta}_{l}$ and $%
\bar{V}\equiv V/2zt$; while the perturbation is
\begin{equation}
\hat{V}=-\sum_{l=1,\sigma }^{2}(\hat{a}_{l\sigma }^{\dag }+\hat{a}_{l\sigma
})\psi _{l+1,\sigma }=\sum_{l=1,\sigma }^{2}\hat{V}_{l\sigma }\psi
_{l+1,\sigma }  \label{2c perturbation}
\end{equation}%
with $\hat{V}_{l\sigma }\equiv -(\hat{a}_{l\sigma }^{\dag }+\hat{a}_{l\sigma
})$.

Similar to the one-component case, the unperturbed energy for a supercell
with $g$ atoms is $E_{g}^{\left( 0\right) }\equiv E_{\left\{ g_{1\uparrow
},g_{1\downarrow };g_{2\uparrow },g_{2\downarrow }\right\} }^{\left(
0\right) }=\min \{E_{n}^{\left( 0\right) }\}_{n=1,2,\ldots }$, which means $%
E_{g}^{\left( 0\right) }<E_{g\pm 1}^{\left( 0\right) }$. Thus, the
zero-order energy is
\begin{eqnarray}
E_{\left\{ g_{1\uparrow },g_{1\downarrow };g_{2\uparrow },g_{2\downarrow
}\right\} }^{\left( 0\right) } &=&\frac{\bar{U}}{2}(g_{1\uparrow
}^{2}+g_{1\downarrow }^{2}+g_{2\uparrow }^{2}+g_{2\downarrow }^{2}-g)+
\notag \\
&&\bar{V}\left( g_{1\uparrow }g_{1\downarrow }+g_{2\uparrow }g_{2\downarrow
}\right) -\left( \bar{\mu}_{\uparrow }g_{\uparrow }+\bar{\mu}_{\downarrow
}g_{\downarrow }\right)  \notag \\
&&+\bar{\Delta}g_{2}+2\left( \psi _{1\uparrow }\psi _{2\uparrow }+\psi
_{1\downarrow }\psi _{2\downarrow }\right) ,
\end{eqnarray}%
and the second-order perturbation energy is
\begin{eqnarray}
E_{g}^{\left( 2\right) } &\equiv &E_{\left\{ g_{1\uparrow },g_{1\downarrow
};g_{2\uparrow },g_{2\downarrow }\right\} }^{\left( 2\right) }=\sum_{n\neq g}%
\frac{|\langle n|\hat{V}|g\rangle |^{2}}{E_{g}^{\left( 0\right)
}-E_{n}^{\left( 0\right) }}  \notag \\
&=&E_{1\uparrow ,g}^{\left( 2\right) }+E_{1\downarrow ,g}^{\left( 2\right)
}+E_{2\uparrow ,g}^{\left( 2\right) }+E_{2\downarrow ,g}^{\left( 2\right) },
\end{eqnarray}%
where ($\bar{\sigma}\equiv -\sigma $)
\begin{widetext}
\begin{subequations}
\begin{eqnarray}
E_{1\sigma ,g}^{\left( 2\right) } &=&\psi _{2\sigma }^{2}\left[ \frac{%
g_{1\sigma }+1}{-\bar{U}g_{1\sigma }-\bar{V}g_{1\bar{\sigma}}+\bar{\mu}%
_{\sigma }}+\frac{g_{1\sigma }}{\bar{U}\left( g_{1\sigma }-1\right) +\bar{V}%
g_{1\bar{\sigma}}-\bar{\mu}_{\sigma }}\right] ,  \label{2c_E2} \\
E_{2\sigma ,g}^{\left( 2\right) } &=&\psi _{1\sigma }^{2}\left[ \frac{%
g_{2\sigma }+1}{-\bar{U}g_{2\sigma }-\bar{V}g_{2\bar{\sigma}}+\bar{\mu}%
_{\sigma }-\bar{\Delta}}+\frac{g_{2\sigma }}{\bar{U}\left( g_{2\sigma
}-1\right) +\bar{V}g_{2\bar{\sigma}}-\bar{\mu}_{\sigma }+\bar{\Delta}}\right]
.
\end{eqnarray}%
\end{subequations}

Hence, the Landau expansion takes the form
\begin{equation}
E=a_{0}+a_{2\uparrow }\psi _{1\uparrow }^{2}+a_{2\downarrow }\psi
_{1\downarrow }^{2}+b_{2\uparrow }\psi _{2\uparrow }^{2}+b_{2\downarrow
}\psi _{2\downarrow }^{2}+c_{2\uparrow }\psi _{1\uparrow }\psi _{2\uparrow
}+c_{2\downarrow }\psi _{1\downarrow }\psi _{2\downarrow }+\mathcal{O}\left(
\psi ^{4}\right) .  \notag
\end{equation}%
The coefficients are
\begin{subequations}
\label{2c_coeffient}
\begin{eqnarray}
a_{0} &=&\frac{\bar{U}}{2}(g_{1\uparrow }^{2}+g_{1\downarrow
}^{2}+g_{2\uparrow }^{2}+g_{2\downarrow }^{2}-g)+\bar{\Delta}g_{2}+\bar{V}%
\left( g_{1\uparrow }g_{1\downarrow }+g_{2\uparrow }g_{2\downarrow }\right)
-\left( \bar{\mu}_{\uparrow }g_{\uparrow }+\bar{\mu}_{\downarrow
}g_{\downarrow }\right) , \\
a_{2\sigma } &=&\frac{g_{2\sigma }+1}{-\bar{U}g_{2\sigma }-\bar{V}g_{2\bar{%
\sigma}}+\bar{\mu}_{\sigma }-\bar{\Delta}}+\frac{g_{2\sigma }}{\bar{U}\left(
g_{2\sigma }-1\right) +\bar{V}g_{2\bar{\sigma}}-\bar{\mu}_{\sigma }+\bar{%
\Delta}}, \\
b_{2\sigma } &=&\frac{g_{1\sigma }+1}{-\bar{U}g_{1\sigma }-\bar{V}g_{1\bar{%
\sigma}}+\bar{\mu}_{\sigma }}+\frac{g_{1\sigma }}{\bar{U}\left( g_{1\sigma
}-1\right) +\bar{V}g_{1\bar{\sigma}}-\bar{\mu}_{\sigma }},\ c_{2\sigma }=2.
\end{eqnarray}
\end{subequations}
To derive the stability condition, we need to consider the following
derivative at the point $E_{\mathbf{0}}\equiv E\left( \psi
_{1\uparrow }=0,\psi _{1\downarrow }=0,\psi _{2\uparrow }=0,\psi
_{2\downarrow }=0\right) $:
\begin{eqnarray*}
\mathcal{D} &=&\left\vert
\begin{array}{cccc}
\frac{\partial ^{2}E_{\mathbf{0}}}{\partial \psi _{1\uparrow }^{2}} & \frac{%
\partial ^{2}E_{\mathbf{0}}}{\partial \psi _{1\uparrow }\partial \psi
_{1\downarrow }} & \frac{\partial ^{2}E_{\mathbf{0}}}{\partial \psi
_{1\uparrow }\partial \psi _{2\uparrow }} & \frac{\partial ^{2}E_{\mathbf{0}}%
}{\partial \psi _{1\uparrow }\partial \psi _{2\downarrow }} \\
\frac{\partial ^{2}E_{\mathbf{0}}}{\partial \psi _{1\downarrow }\partial
\psi _{1\uparrow }} & \frac{\partial ^{2}E_{\mathbf{0}}}{\partial \psi
_{1\downarrow }^{2}} & \frac{\partial ^{2}E_{\mathbf{0}}}{\partial \psi
_{1\downarrow }\partial \psi _{2\uparrow }} & \frac{\partial ^{2}E_{\mathbf{0%
}}}{\partial \psi _{1\downarrow }\partial \psi _{2\downarrow }} \\
\frac{\partial ^{2}E_{\mathbf{0}}}{\partial \psi _{2\uparrow }\partial \psi
_{1\uparrow }} & \frac{\partial ^{2}E_{\mathbf{0}}}{\partial \psi
_{2\uparrow }\partial \psi _{1\downarrow }} & \frac{\partial ^{2}E_{\mathbf{0%
}}}{\partial \psi _{2\uparrow }^{2}} & \frac{\partial ^{2}E_{\mathbf{0}}}{%
\partial \psi _{2\uparrow }\partial \psi _{2\downarrow }} \\
\frac{\partial ^{2}E_{\mathbf{0}}}{\partial \psi _{2\downarrow }\partial
\psi _{1\uparrow }} & \frac{\partial ^{2}E_{\mathbf{0}}}{\partial \psi
_{2\downarrow }\partial \psi _{1\downarrow }} & \frac{\partial ^{2}E_{%
\mathbf{0}}}{\partial \psi _{2\downarrow }\partial \psi _{2\uparrow }} &
\frac{\partial ^{2}E_{\mathbf{0}}}{\partial \psi _{2\downarrow }^{2}}%
\end{array}%
\right\vert =\left\vert
\begin{array}{cccc}
2a_{2\uparrow } & 0 & c_{2\uparrow } & 0 \\
0 & 2a_{2\downarrow } & 0 & c_{2\downarrow } \\
c_{2\uparrow } & 0 & 2b_{2\uparrow } & 0 \\
0 & c_{2\downarrow } & 0 & 2b_{2\downarrow }%
\end{array}%
\right\vert  \\
&=&c_{2\uparrow }^{2}c_{2\downarrow }^{2}-4a_{2\downarrow }b_{2\downarrow
}c_{2\uparrow }^{2}-4a_{2\uparrow }b_{2\uparrow }c_{2\downarrow
}^{2}+16a_{2\uparrow }a_{2\downarrow }b_{2\uparrow }b_{2\downarrow }.
\end{eqnarray*}
\end{widetext}
When $\mathcal{D}=0$, the insulating phase is no longer stable, thus a
phase\ transition takes place. This critical\ condition can be written as $%
4a_{2\uparrow }b_{2\uparrow }=c_{2\uparrow }^{2}$ or $4a_{2\downarrow
}b_{2\downarrow }=c_{2\downarrow }^{2}$. (Either one being satisfied will
destabilize the system.) As a result, there exists two phase boundaries,
each one for a spin component, enclosing a region in the phase diagram where
one species is in the superfluid phase while the other one is still in the
Mott insulating state. The phase boundaries for each can be written as ($%
\tilde{V}\equiv V/U$)%
\begin{widetext}
\begin{equation}
\tilde{t}_{c\sigma }=\frac{1}{2}\sqrt{\frac{[(g_{1\sigma }-1)-\tilde{\mu}%
_{\sigma }+\tilde{V}g_{1\bar{\sigma}}](-g_{1\sigma }+\tilde{\mu}_{\sigma }-%
\tilde{V}g_{1\bar{\sigma}})[(g_{2\sigma }-1)-\tilde{\mu}_{\sigma }+\tilde{%
\Delta}+\tilde{V}g_{2\bar{\sigma}}](-g_{2\sigma }+\tilde{\mu}_{\sigma }-%
\tilde{\Delta}-\tilde{V}g_{2\bar{\sigma}})}{(\tilde{\mu}_{\sigma }+1-\tilde{V%
}g_{1\bar{\sigma}})(\tilde{\mu}_{\sigma }+1-\tilde{\Delta}-\tilde{V}g_{2\bar{%
\sigma}})}}.  \label{2c critical}
\end{equation}%
\end{widetext}

\subsection{Phase diagrams}

Similar to the single component case, we first determine the unperturbed
groundstate and the range of the chemical potential, then depict the
critical curve Eq. (\ref{2c critical}). As $\tilde{V}\equiv V/U$ varies,
there exists various filling configurations for the Mott state (unperturbed
state). For the sake of clarity, we focus on the\ case $\bar{V}=1$, which is
most accessible in current experiments.

When the inter-species and intra-species repulsions are identical, the
filling configuration for groundstate is very like the single component
case, but there are much complicated spin-texture pattern, the spin
imbalance $m\equiv g_{\uparrow }-g_{\downarrow }$, (something like SDW)
besides the CDW-like pattern in the former case. We list all the filling
patterns for one spin component in the Appendix.

After comparing $E_{g}^{\left( 0\right) }<E_{g+1}^{\left( 0\right) },\
E_{g}^{\left( 0\right) }<E_{g-1}^{\left( 0\right) }$, we find the
constraints on chemical potential $\tilde{\mu}_{\sigma }$. For the most
imbalanced case $\left\{ g;0\right\} $, $\tilde{\mu}_{\sigma }\in \left(
g-1,g\right) $. For other filling patterns, the results are listed below.
\begin{widetext}
\begin{equation*}
\begin{tabular}{|c|c|c|}
\hline $\tilde{\Delta}$ & $g\in \mathrm{even}$ & $g\in \mathrm{odd}$
\\ \hline $\left( p-1,p\right) $ & $\tilde{\mu}_{\sigma }\in \left(
\frac{1}{2}\left[
g+\left( p-2\right) \right] ,\frac{1}{2}\left( g-p\right) +\tilde{\Delta}%
\right) $ & $\tilde{\mu}_{\sigma }\in \left( \frac{1}{2}\left[
g-\left(
p+1\right) \right] +\tilde{\Delta},\frac{1}{2}\left[ g+\left( p-1\right) %
\right] \right) $ \\ \hline $\left( p,p+1\right) $ &
$\tilde{\mu}_{\sigma }\in \left( \frac{1}{2}\left[ g-\left(
p+2\right) \right] +\tilde{\Delta},\frac{1}{2}\left( g+p\right)
\right) $ & $\tilde{\mu}_{\sigma }\in \left( \frac{1}{2}\left[
g+\left(
p-1\right) \right] ,\frac{1}{2}\left[ g-\left( p+1\right) \right] +\tilde{%
\Delta}\right) $ \\ \hline
\end{tabular}%
\end{equation*}
\end{widetext}

\begin{figure}[tbp]
\begin{center}
\includegraphics[width=0.5\textwidth]{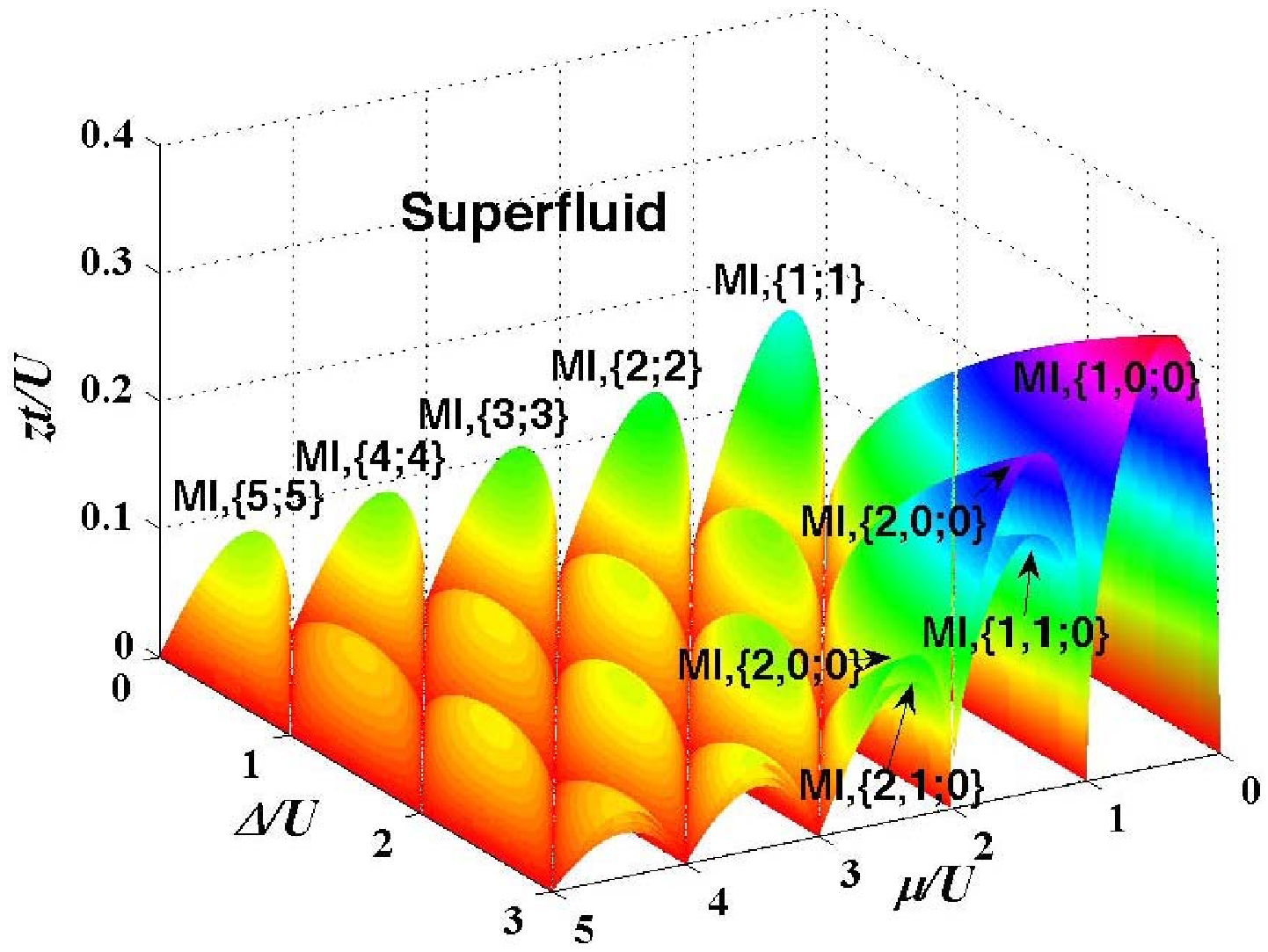}
\end{center}
\caption{(color online) The phase diagram spin-$\uparrow $ atoms in
two-component Bose-Hubbard model when $V=U$. MI stands for Mott insulator. $%
\left\{ g_{1\uparrow },g_{1\downarrow };g_{2\uparrow },g_{2\downarrow
}\right\} $ reflects the filling configuration in Mott state, where $g_{l%
\protect\sigma }$ particles with spin-$\protect\sigma $ are located in the $%
l $-th site in a supercell.}
\label{2c_phase_diagram_3D}
\end{figure}

\begin{figure}[tbp]
\begin{center}
\includegraphics[width=0.52\textwidth]{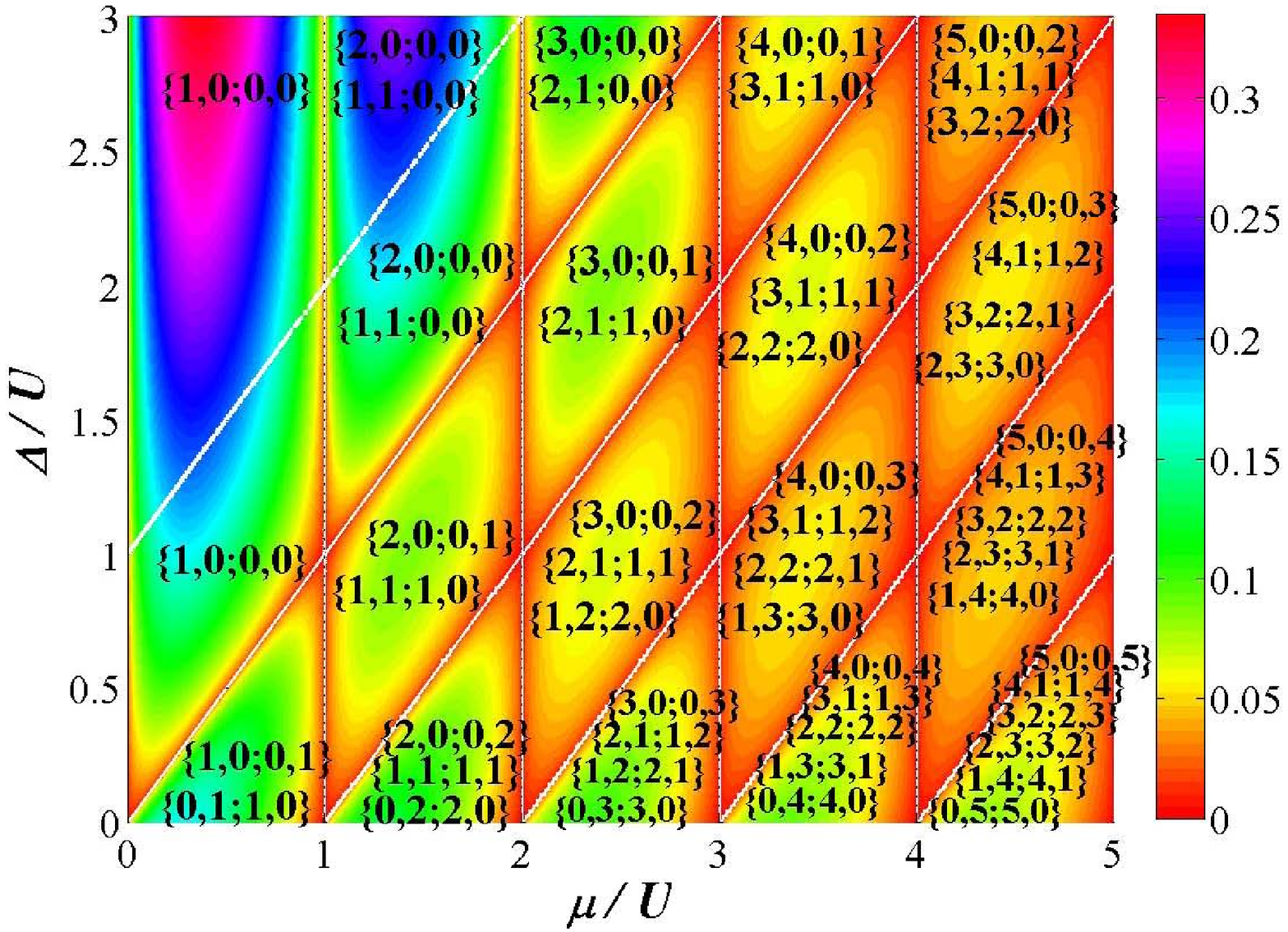}
\end{center}
\caption{(color online) The contour diagram for two-component case when $V=U$%
, revealing different CDW and SDW orders (spin imbalance).}
\label{2c_phase_diagram_2D}
\end{figure}

With these, we plot the phase diagram for spin-$\uparrow $ atoms in Figs. (%
\ref{2c_phase_diagram_3D}) and (\ref{2c_phase_diagram_2D}). The global
structure of these diagrams are similar to the one-component case, but there
exist many small lobes hiding in a large lobe, as shown in Fig. (\ref%
{2c_phase_diagram_3D}). They are the phase boundaries of various filling
patterns in the Mott phase. The details are listed in Fig. (\ref%
{2c_phase_diagram_2D}), where the sequence reflects the actual position of
lobes in Figs. (\ref{2c_phase_diagram_3D}). For example, $\left\{
1,0;0,1\right\} $ is written above $\left\{ 0,1;1,0\right\} $, this means
that in the 3D phase diagram, the lobe of filling $\left\{ 1,0;0,1\right\} $
is also located beyond the lobe of $\left\{ 0,1;1,0\right\} $. The
illustrate these multiple lobes more clearly, in Fig. (\ref%
{2c_phase_diagram_crosssection}) we draw the cross-sections of the 3D phase
diagram, $\Delta =0$ (a) and $\Delta =U$ (b). The structures of other
cross-sections are similar to these two. Every lobes from bottom to top
corresponds to a filling configuration in Fig. (\ref{2c_phase_diagram_2D}).
From them, we find that there may exist weak superfluid (WSF) regions
between two lobes that have the same particle number $g$. In WSF regions,
one spin species becomes superfluid while the other type remains insulating.

\begin{figure}[tbp]
\begin{center}
\includegraphics[width=0.5\textwidth]{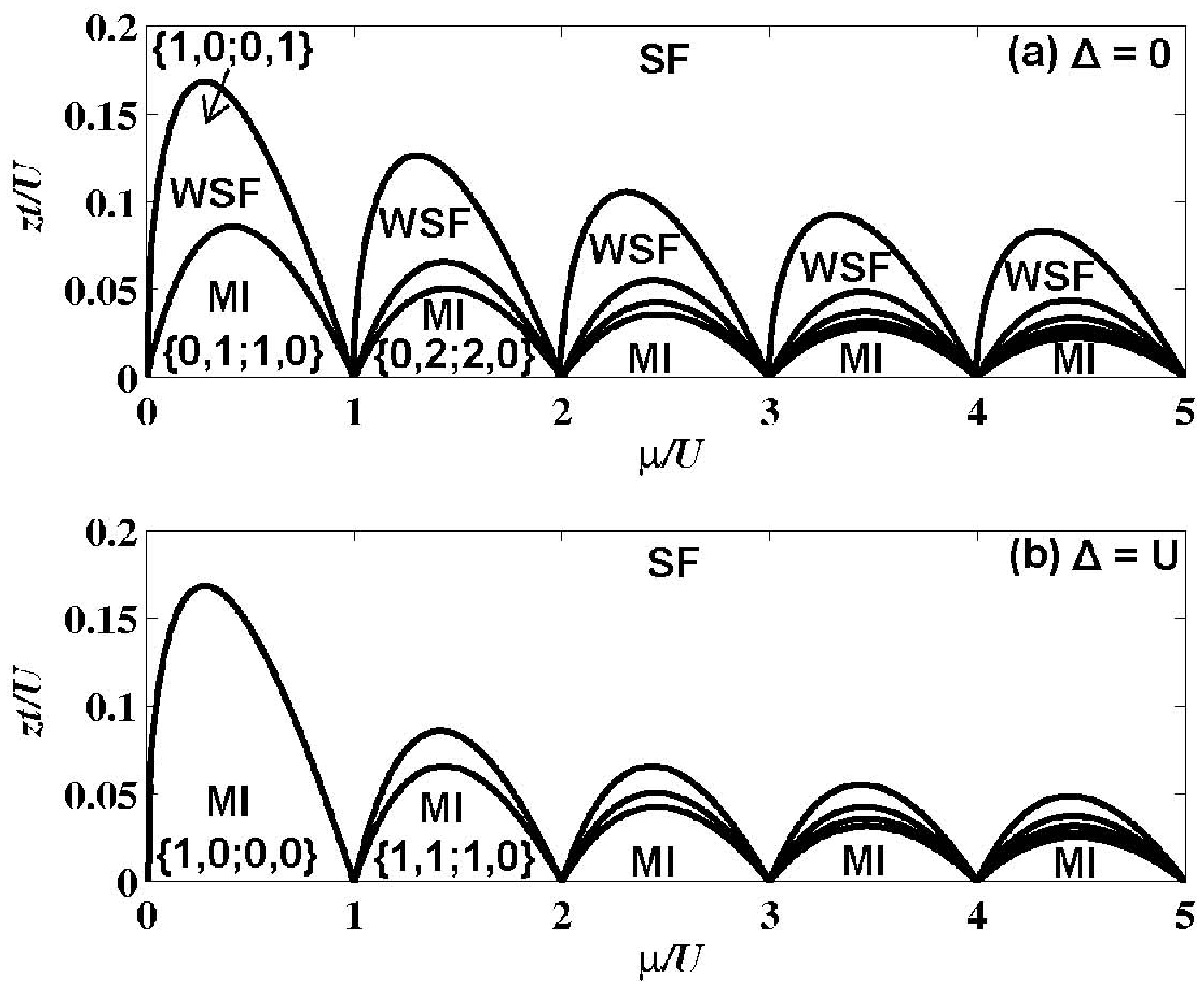}
\end{center}
\caption{The cross-section for two-component case when $V=U$, showing the
multiple lopes and possible weak superfluid regions (WSF). SF stands for
superfluid. Detailed filling conditions can be found in Fig. (\protect\ref%
{2c_phase_diagram_2D}).}
\label{2c_phase_diagram_crosssection}
\end{figure}

\subsection{Spin dynamics in biased superlattices}

In addition to the phase transitions that we describe above, in the Mott
insulating phase, the two-component BEC trapped biased superlattices can
exhibit spin dynamics and SDW-like patterns at different filling
configurations.

At unit filling, Hubbard model [Eq. (\ref{2c BH})] can be transformed into a
Heisenberg-type model, via a canonical transformation and a second-order
perturbation over $t$\cite{Barthel}, namely,
\begin{equation}
\hat{H}_{\mathrm{eff}}=-J\sum_{\left\langle ij\right\rangle }(\hat{S}_{i}^{x}%
\hat{S}_{j}^{y}+\hat{S}_{i}^{y}\hat{S}_{j}^{y})+\left( J-J_{s}\right)
\sum_{\left\langle ij\right\rangle }\hat{S}_{i}^{z}\hat{S}_{j}^{z},
\label{H-}
\end{equation}%
where the (pseudo) spin operator is defined as usual, $\hat{S}_{i}^{\alpha }=%
\frac{1}{2}\sum_{\mu \nu }\hat{a}_{i\mu }^{\dag }\sigma _{\mu \nu }^{\alpha }%
\hat{a}_{i\nu }$ with $\sigma ^{\alpha }$ the Pauli matrix, $\alpha =x,y,z$.
The effective exchange energy is
\begin{equation}
J=\frac{4t^{2}V}{V^{2}-\Delta ^{2}},\ J_{s}=\frac{8t^{2}U}{U^{2}-\Delta ^{2}}%
.
\end{equation}

When $V\ll U$ and $V\gg U$, $J=0$, then $\hat{H}_{\mathrm{eff}%
}=-J_{s}\sum_{\left\langle ij\right\rangle }S_{i}^{z}S_{j}^{z}$, which is an
Ising model. There is no spin wave excitations in this case.

When $V=U$, $J_{s}=2J$, then $H_{\mathrm{eff}}=-J\sum_{\left\langle
ij\right\rangle }\mathbf{S}_{i}\cdot \mathbf{S}_{j}$, which is a Heisenberg
model. For ferromagnetic case ($\Delta <U$, $J>0$), we define
Holstein-Primakoff (HP) transformation as
\begin{equation}
\hat{S}^{+}=\sqrt{2S-\hat{c}^{\dag }\hat{c}}\,\hat{c},\ \hat{S}^{-}=\hat{c}%
^{\dag }\sqrt{2S-\hat{c}^{\dag }\hat{c}},\ \hat{S}^{z}=S-\hat{c}^{\dag }\hat{%
c},
\end{equation}%
where bosonic operator $\hat{c}^{\dag }\left( \hat{c}\right) $ creates
(annihilates) spin deviations, $S^{\pm }=S^{x}\pm iS^{y}$,$\ S$ is the total
spin. For low excited states, spin deviation is small, thus $\sqrt{2S-\hat{c}%
^{\dag }\hat{c}}\approx \sqrt{2S}$; applying this transformation up to
second order, the ferromagnetic Heisenberg Hamiltonian becomes
\begin{eqnarray}
\hat{H}_{\mathrm{eff}} &=&-J\sum_{\left\langle ij\right\rangle }\mathbf{\hat{%
S}}_{i}\cdot \mathbf{\hat{S}}_{j}  \notag \\
&=&E_{0}+2zJS\sum_{i}\hat{c}_{i}^{\dag }\hat{c}_{i}-JS\sum_{\left\langle
ij\right\rangle }(\hat{c}_{i}^{\dag }\hat{c}_{j}+h.c.).
\end{eqnarray}%
Here the ground state energy is $E_{0}=-NzJS^{2}$ with $N$ the number of
lattice site. This Hamiltonian is easy to diagnolize, the excitation energy
in square lattice (in arbitrary dimensions) is
\begin{equation}
\epsilon _{k}=2JS[z-\sum_{\mathbf{\delta }}\cos \left( \mathbf{k}\cdot
\mathbf{\delta }\right) ]\approx JS\sum_{\mathbf{\delta }}\left( \mathbf{k}%
\cdot \mathbf{\delta }\right) ^{2}=2JSk^{2}.
\end{equation}%
Here, we set the lattice spacing to unity.

For anti-ferromagnetic case $\Delta \gtrsim U$, $J<0$, the
single-occupied state is \emph{metastable} and it would decay to a
triplet groundstate which satisfies the Libe-Mattis
theorem\cite{Lieb}. However, this metastable state can be
prepared\cite{Demler} and probed\cite{Rosch} in an
experiment-accessible time interval\cite{Bloch}, if the band width
of a single particle $t$\ is much smaller than its band gap $\Delta
E=\Delta -U$, $t\ll \Delta E$\cite{Rosch}. Under this circumstance,
we can perform the HP transformation in $A$-$B$ sublattices and find
the long-wave excitation in square lattices is
\begin{equation}
\epsilon _{k}=2z\left\vert J\right\vert S\sqrt{1-\gamma _{k}^{2}}\approx
\sqrt{2z}\left\vert J\right\vert Sk,  \label{dispersion}
\end{equation}%
where $\gamma _{k}=z^{-1}\sum_{\mathbf{\delta }}\cos \left( \mathbf{k}\cdot
\mathbf{\delta }\right) $.

\section{Conclusions}

In this paper, using a decoupling approximation, we analyze the possible
phase diagrams of one-and two-component Bose-Hubbard models in optical
superlattices in the mean-field level. As the potential bias $\Delta $ of
the superlattice, the atomic repulsion $U$ and hopping $t$, the filling
configuration and chemical potential $\mu $ varies, we discover complex
phases in different parameter regions. For one-component case, there exists
Mott states with CDW order and corresponding MI-SF transitions. For
two-component case, besides the CDW in Mott state, there also exists weak SF
regions where one spin component holds the CDW order while the other
component becoming superfluid. In addition, the spin imbalance for a certain
filling configuration (see Appendix) implies the existence of an SDW-like
order in the Mott state. We also calculate spin dynamics of the
two-component model at unit filling. The results explicit different
low-energy dispersions for different $\Delta /U$. The features can be tested
via many accessible probing techniques\cite{Bloch1,Iskin1} in current
experiments with single-site resolution\cite{Greiner1}.

\begin{acknowledgments}
This research is supported by NCET, NFSC Grant No. 10874017.
\end{acknowledgments}

\appendix{}

\section{Filling Configuration for one-component case}

The groundstate filling configurations $\left\{ g_{1};g_{2}\right\} $ for
different $g$ (total number of particles in a supercell) and $\tilde{\Delta}%
\equiv \Delta /U$ are listed in the following table. Here, $p$ is an even
number, $p-1\geq 0$, reflecting the ratio between potential bias $\Delta $
and inter-atomic repulsion $U$. The notation $\left\{ g_{1};g_{2}\right\} $
means that there are $g_{1}$ particles in the left (deep) site and $g_{2}$
particles in the right (shallow) site in a supercell.
\begin{widetext}
\begin{equation*}
\begin{tabular}{|c|c|c|}
\hline
$\tilde{\Delta}$ & $g\in \mathrm{even}$ & $g\in \mathrm{odd}$ \\ \hline
\multicolumn{1}{|l|}{$\left( p-1,p\right) $} & $%
\begin{array}{l}
\left\{ \frac{g}{2}+\frac{p}{2};\frac{g}{2}-\frac{p}{2}\right\} \\
\multicolumn{1}{c}{\left\{ g,0\right\} ,\text{ }\mathrm{if}\text{ }g\leq p}%
\end{array}%
$ & $%
\begin{array}{c}
\left\{ \frac{g+1}{2}+\left( \frac{p}{2}-1\right) ;\frac{g-1}{2}-\left(
\frac{p}{2}-1\right) \right\} \\
\left\{ g,0\right\} ,\text{ }\mathrm{if}\text{ }g\leq p-1%
\end{array}%
$ \\ \hline
\multicolumn{1}{|l|}{$\left( p,p+1\right) $} & $%
\begin{array}{l}
\left\{ \frac{g}{2}+\frac{p}{2};\frac{g}{2}-\frac{p}{2}\right\} \\
\multicolumn{1}{c}{\left\{ g,0\right\} ,\text{ }\mathrm{if}\text{ }g\leq p}%
\end{array}%
$ & $%
\begin{array}{l}
\left\{ \frac{g+1}{2}+\frac{p}{2};\frac{g-1}{2}-\frac{p}{2}\right\} \\
\multicolumn{1}{c}{\left\{ g,0\right\} ,\text{ }\mathrm{if}\text{ }g\leq p+1}%
\end{array}%
$ \\ \hline
\end{tabular}%
\end{equation*}
\end{widetext}

\section{Filling Configuration for two-component case}

The filling configurations $\left\{ g_{1\uparrow },g_{1\downarrow
};g_{2\uparrow },g_{2\downarrow }\right\} $ for two-component case is listed
in the following tables. Here, $p$ is still an even number that set the
value of $\tilde{\Delta}$, $k,l,q$ are integers that take values in certain
intervals. The notation $\left\{ g_{1\uparrow },g_{1\downarrow
};g_{2\uparrow },g_{2\downarrow }\right\} $ means that there are $%
g_{1\uparrow }$ spin-$\uparrow $ atoms and $g_{1\downarrow }$ spin-$%
\downarrow $ atoms in the left (deep) site, $g_{2\uparrow }$ spin-$\uparrow $
atoms and $g_{2\downarrow }$ spin-$\downarrow $ atoms in the right (shallow)
site in a supercell.
\begin{widetext}
\begin{equation*}
\begin{tabular}{|c|c|}
\hline $\tilde{\Delta}$ & $g\in \mathrm{even}$ \\ \hline
\multicolumn{1}{|l|}{$\left( p-1,p\right) $} & $%
\begin{array}{c}
\left\{ \frac{g}{2}+k-l-q,\frac{p}{2}-\left( k-l\right) +q;l+q,\frac{g}{2}-%
\frac{p}{2}-l-q\right\}  \\
k\in \left[ 0,\frac{g}{2}\right] ,l\in \left[ 0,k\right] ,q\in \left[ 0,%
\frac{g}{2}-\frac{p}{2}-l\right]
\end{array}%
$ \\ \hline
\multicolumn{1}{|l|}{$\left( p,p+1\right) $} & $%
\begin{array}{c}
\left\{ \frac{g}{2}+k-l-q,\frac{p}{2}-\left( k-l\right) +q;l+q,\frac{g}{2}-%
\frac{p}{2}-l-q\right\}  \\
k\in \left[ 0,\frac{g}{2}\right] ,l\in \left[ 0,k\right] ,q\in \left[ 0,%
\frac{g}{2}-\frac{p}{2}-l\right]
\end{array}%
$ \\ \hline
\end{tabular}%
\end{equation*}%
\begin{equation*}
\begin{tabular}{|c|c|}
\hline $\tilde{\Delta}$ & $g\in \mathrm{odd}$ \\ \hline
\multicolumn{1}{|l|}{$\left( p-1,p\right) $} & $%
\begin{array}{c}
\left\{ \frac{g+1}{2}+\left( k-l\right) -q,\left( \frac{p}{2}-1\right)
-\left( k-l\right) +q;l+q,\frac{g-1}{2}-\left( \frac{p}{2}-1\right) -\left(
l+q\right) \right\}  \\
k\in \left[ 0,\frac{g-1}{2}\right] ,l\in \left[ 0,k\right] ,q\in \left[ 0,%
\frac{g+1}{2}-\frac{p}{2}-l\right]
\end{array}%
$ \\ \hline
\multicolumn{1}{|l|}{$\left( p,p+1\right) $} & $%
\begin{array}{c}
\left\{ \frac{g+1}{2}+k-l-q,\frac{p}{2}-\left( k-l\right) +q;l+q,\frac{g-1}{2%
}-\frac{p}{2}-l-q\right\}  \\
k\in \left[ 0,\frac{g-1}{2}\right] ,l\in \left[ 0,k\right] ,q\in \left[ 0,%
\frac{g-1}{2}-\frac{p}{2}-l\right]
\end{array}%
$ \\ \hline
\end{tabular}%
\end{equation*}

We can also define the spin imbalance $\left( m_{1}\equiv g_{1\uparrow
}-g_{1\downarrow },m_{2}\equiv g_{2\uparrow }-g_{2\downarrow }\right) $ for
a given filling pattern, the results are listed in the following table.
\begin{equation*}
\begin{tabular}{|c|c|c|}
\hline $\tilde{\Delta}$ & $g\in \mathrm{even}$ & $g\in \mathrm{odd}$
\\ \hline
\multicolumn{1}{|l|}{$\left( p-1,p\right) $} & $m_{1}=\frac{g}{2}-\frac{p}{2}%
+2\left( k-l-q\right) ,m_{2}=2\left( l+q\right) -\left( \frac{g}{2}-\frac{p}{%
2}\right) $ & $m_{1}=\frac{g+3}{2}-\frac{p}{2}+2\left( k-l-q\right)
,m_{2}=2\left( l+q\right) -\left( \frac{g+1}{2}-\frac{p}{2}\right) $ \\
\hline
\multicolumn{1}{|l|}{$\left( p,p+1\right) $} & $m_{1}=\frac{g}{2}-\frac{p}{2}%
+2\left( k-l-q\right) ,m_{2}=2\left( l+q\right) -\left( \frac{g}{2}-\frac{p}{%
2}\right) $ & $m_{1}=\frac{g+1}{2}-\frac{p}{2}+2\left( k-l-q\right)
,m_{2}=2\left( l+q\right) -\left( \frac{g-1}{2}-\frac{p}{2}\right) $ \\
\hline
\end{tabular}%
\end{equation*}

For filling configuration $\left\{ g;0\right\} $, we combine both filling
pattern and spin imbalance in the following.
\begin{equation*}
\begin{tabular}{|c|c|c|}
\hline $\tilde{\Delta}$ & $g\in \mathrm{even}$ & $g\in \mathrm{odd}$
\\ \hline
\multicolumn{1}{|l|}{$\left( p-1,p\right) $} & $%
\begin{array}{c}
\left\{ \frac{g}{2}+k,\frac{g}{2}-k;0,0\right\} ,g\leq p,k\in \left[ 0,\frac{%
g}{2}\right]  \\
m_{1}=2k,m_{2}=0%
\end{array}%
$ & $%
\begin{array}{c}
\left\{ \frac{g+1}{2}+k,\frac{g-1}{2}-k;0,0\right\} ,g\leq p-1,k\in \left[ 0,%
\frac{g-1}{2}\right]  \\
m_{1}=2k+1,m_{2}=0%
\end{array}%
$ \\ \hline
\multicolumn{1}{|l|}{$\left( p,p+1\right) $} & $%
\begin{array}{c}
\left\{ \frac{g}{2}+k,\frac{g}{2}-k;0,0\right\} ,g\leq p,k\in \left[ 0,\frac{%
g}{2}\right]  \\
m_{1}=2k,m_{2}=0%
\end{array}%
$ & $%
\begin{array}{c}
\left\{ \frac{g+1}{2}+k,\frac{g-1}{2}-k;0,0\right\} ,g\leq p+1,k\in \left[ 0,%
\frac{g-1}{2}\right]  \\
m_{1}=2k+1,m_{2}=0%
\end{array}%
$ \\ \hline
\end{tabular}%
\end{equation*}
\end{widetext}

\end{document}